\documentclass[doublecol]{epl2}

\title{Point-Contact Spectroscopy of Iron-Based Layered Superconductor LaO$_{0.9}$F$_{0.1-\delta}$FeAs}
\shorttitle{Title} 

\author{Lei Shan\thanks{E-mail: \email{lshan@aphy.iphy.ac.cn}} \and Yonglei Wang \and Xiyu Zhu \and Gang Mu \and Lei Fang \and Cong Ren \and Hai-Hu Wen}
\shortauthor{F. Author \etal}

\institute{National Laboratory for Superconductivity, Institute of Physics $\&$ Beijing National
Laboratory for Condensed Matter Physics, Chinese Academy of Sciences, P.O. Box 603, Beijing 100190,
China
} 

\abstract{ We present point-contact spectroscopy data for junctions between a normal metal and the
newly discovered F-doped superconductor LaO$_{0.9}$F$_{0.1-\delta}$FeAs (F-LaOFeAs). A zero-bias
conductance peak was observed and its shape and magnitude suggests the presence of Andreev bound
states at the surface of F-LaOFeAs, which provides a possible evidence of an unconventional pairing
symmetry with a nodal gap function. The maximum gap value $\Delta_0\approx3.9\pm0.7$meV was
determined from the measured spectra, in good agreement with the recent experiments on specific
heat and lower critical field.}

\begin{document}

\maketitle

\section{Introduction}
Copper-based layered superconductors have attracted extensive attention due to their high
superconducting transition temperature ($T_c$) and the underlying rich physics of strong electron
correlation. Although $d_{x^2-y^2}$ pairing symmetry has been proved in cuprates
\cite{Tsuei2000RMP}, the mechanism of high temperature superconductivity is still not settled down.
In order to open a new path to clarify this issue, much efforts have been devoted to seeking new
transition-metal-based superconductors other than cuprates. Recently, superconductivity was
observed in the Fe-based layered material La[O$_{1-x}$F$_x$]FeAs ($x = 0.05\sim0.12$) with
transition temperatures $T_c\approx26$K \cite{Kamihara2008JACS}. Immediately after this exciting
discovery, Ren {\it et al.} reported that the $T_c$ could reach 55 K in Sm[O$_{1-x}$F$_x$]FeAs
\cite{Ren2008cond2}. Superconductivity at 25K was also found in Sr-substituted samples
La$_{1-x}$Sr$_x$OFeAs without F-doping \cite{WenHH2008EPL}. This provides a new chance to
understand high-$T_c$ superconductivity in non-cuprate systems. Recent experiments indicated that
La[O$_{1-x}$F$_x$]FeAs has a low carrier density \cite{Chen2008cond,Zhu2008cond} with strong
electron-electron correlation \cite{Athena2008cond}, and its superconducting regime occurs in close
proximity to a long-range ordered antiferromagnetic ground state \cite{Cruz2008cond}. All these
properties are much like that of high-$T_c$ copper oxides and hence a novel superconductivity is
anticipated. Theoretically, unconventional superconductivity beyond $s$-wave pairing was predicted
for Fe-based superconductors by recent calculations
\cite{Boeri2008cond,Mazin2008cond,Cao2008cond,Kuroki2008cond,Dai2008cond,Si2008cond}. Most
interestingly, a nonlinear magnetic field dependence of the electronic specific heat coefficient
$\gamma(H)$ has been found for LaO$_{0.9}$F$_{0.1-\delta}$FeAs (F-LaOFeAs) in the low temperature
limit, indicative of an unconventional pairing symmetry with a nodal gap function
\cite{Mu2008cond}, which is in good agreement with the linear temperature dependence of lower
critical field ($H_{c1}$) observed on the same material \cite{RenC2008cond}. However, the
possibility of novel pairing symmetry in this new superconductor needs to be verified by more
experiments especially by phase-related measurements.

In this letter, we report the study of point-contact spectroscopy on the superconducting F-LaOFeAs
as a function of both temperature and magnetic field. We observed a distinct zero-bias conductance
peak in the conductance-voltage characteristics of the point-contact junctions, suggesting that
F-LaOFeAs has an unconventional pairing symmetry with a nodal gap. The determined maximum gap value
is consistent with the results of specific heat and lower critical field.

\section{Experiment}
The polycrystalline samples of F-LaOFeAs used here were synthesized by using a two-step solid state
reaction method. The superconducting transition temperature $T_c$ defined as the onset of the drop
in resistivity was 27 K with a transition width of $\Delta T_c \approx 3$ K (10\%-90\% of normal
state resistivity). The detailed information about the synthesization is elaborated in a recent
paper \cite{Mu2008cond}. The point-contact junctions are prepared by carefully driving the Pt/ Ir
alloy or Au tips towards the sample surface which is polished by fine sand paper and cleaned by
ultrasonic beforehand. The tip's preparation and the details of the experimental setup were
described elsewhere \cite{Shan2003PRB}. Typical four-terminal and lock-in techniques were used to
measure the conductance-voltage ($dI/dV-V$ or $G-V$) characteristics. Each measurement is comprised
of two successive cycles, to check the absence of heating-hysteresis effects.

\begin{figure}
\onefigure[scale=1.15]{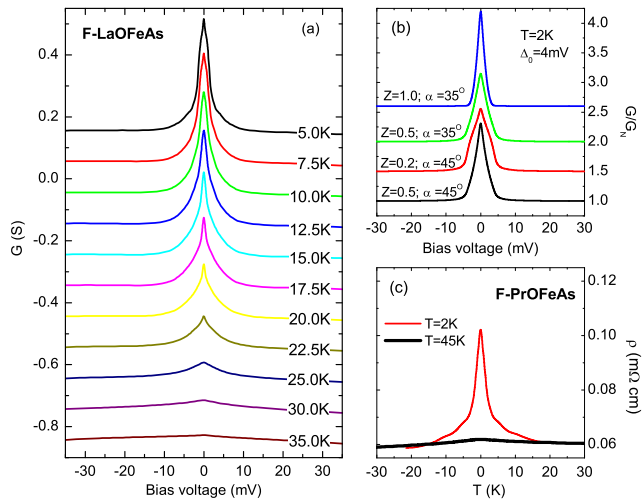} \caption{(a) Conductance-voltage ($G-V$) characteristics of
F-LaOFeAs measured at various temperatures up to above $T_c$ for $H=0T$. All the curves except the
top one are offset downwards for clarity. (b) Calculated spectra for a nodal superconductor
according to the extended BTK model described in text. All the curves except the lowest one are
offset upwards for clarity. (c) $G-V$ curves of F-PrOFeAs measured at 2k and 45K (above $T_c$),
respectively.} \label{fig:fig1}
\end{figure}

\section{Results and discussion}
As shown in Fig.~\ref{fig:fig1}(a), Fig.~\ref{fig:fig3}(a) (raw data) and Fig.~\ref{fig:fig4}(d) or
(e) (normalized spectra), most of the measured spectra have a zero-bias conductance peak (ZBCP)
though the magnitude and shape of the peak varies from one position to another on the sample
surface. The approximate percentage of the occurrence of ZBCP is more than 50\% for the sample we
have studied. Many other spectra are featureless and thus maybe come from the non-superconducting
region on the sample surface (the superconducting volume is above 80\% for the bulk of the sample).
Moreover, such ZBCP was observed in the spectra with various junction resistances from 5 $\Omega$
to 80 $\Omega$ and is not related to the effect of large contact spot. Blonder {\it et al.}
\cite{Blonder1982PRB} have proposed a simplified theory for the $G-V$ curves of an s-wave
superconductor/normal metal junction separated by a barrier of arbitrary strength. The barrier
strength is parametrized by a dimensionless number $Z$ which describes the crossover from metallic
to ideal tunnel junction behavior by $Z=0$ to $Z=\infty$. It was predicted that no sharp ZBCP could
be observed for an s-wave superconductor/normal metal junction with any barrier strength,
furthermore, the normalized conductance ($G/G_N$) at zero bias can not exceed 2 ( the upper limit
of the ideal Andreev reflection). This is obviously not the case of our measurements, in which a
distinct ZBCP was observed and its magnitude can exceed 2. However, this ZBCP can be reasonably
associated to the surface Andreev bound states and is one of the unique features of a nodal
superconductor \cite{Hu1994PRL}, as demonstrated in the cuprate superconductors such as YBCO and
LSCO \cite{Deutscher2005RMP}.

Tanaka {\it et al.} \cite{Tanaka1995PRL,Kashiwaya1996PRB,Kashiwaya2000RPP} extended the BTK model
to deal with the issue of an anisotropic $d$-wave superconductor. In this case, another parameter
$\alpha$ (the angle between the quasiparticle injecting direction and the main crystalline axis)
was introduced in addition to the barrier strength $Z$. Moreover, the isotropic superconducting gap
$\Delta$ in the s-wave BTK model was replaced by an anisotropic gap with $d_{x^2-y^2}$ symmetry,
i.e., $\Delta=\Delta_0cos(2\theta)$ in which $\Delta_0$ is the maximum gap. It was thus predicted
that, for $Z>0$ the ZBCP is formed for all directions in the $a-b$ plane except when tunneling into
the (100) and (010) planes. This ZBCP will be suppressed only when the quasiparticle scattering is
strong enough \cite{Aprili1998PRBr}. The calculated curves using this extended BTK model are
presented in Fig.~\ref{fig:fig1}(b). A ZBCP can be clearly seen in all curves for different
$\alpha$ and $Z$ though its shape and magnitude varies from one to another, which is qualitatively
consistent with the measured spectra in this work.

\begin{figure}
\onefigure[scale=1.3]{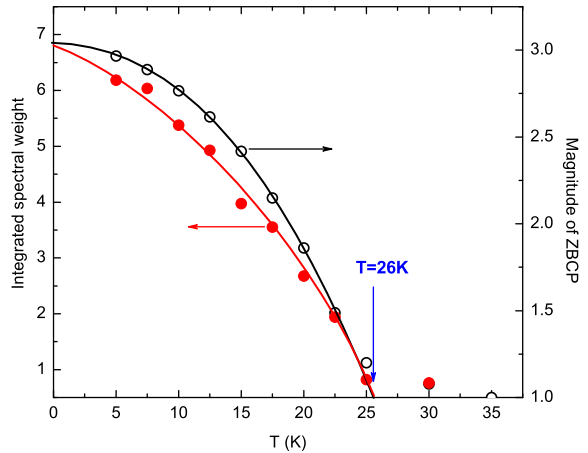} \caption{Temperature dependence of the magnitude of ZBCP and the
integrated spectral weight. Both parameters are obtained from the spectra shown in
Fig.~\ref{fig:fig1}(b) after normalization by the data of 35K.} \label{fig:fig2}
\end{figure}

It is necessary to consider other possible origins of the ZBCP observed here before it can be
ascribed to the surface Andreev bound states of a nodal superconductor. First, a remarkable ZBCP
often shows up if inter-grain Josephson-coupling effect is in series with the point contact
\cite{Shan2003PRB}, or if the role of critical current becomes dominant due to significant
dissipation at the point-contact microconstriction \cite{Sheet2004PRB}. However, these conjectures
also favor two sharp dips besides the central ZBCP which were not observed here. Another mechanism
for the ZBCP is related to magnetic and Kondo scattering coming from the magnetic impurities in or
near the barrier \cite{Appelbaum1966PRL}. However, a ZBCP due to magnetic impurities should be
uncorrelated with the occurrence of superconductivity. In contrast, as shown in
Fig.~\ref{fig:fig2}, both the magnitude and the integrated spectral weight of ZBCP in the present
experiment were found to build up rapidly just below 26K, very close to the bulk $T_c$. In
addition, the magnitude of the ZBCP caused by magnetic impurities should depend on temperature
logarithmically \cite{Appelbaum1966PRL}, which is also inconsistent with our data. Furthermore, as
presented in Fig.~\ref{fig:fig1}(c), similar ZBCP has been observed on another newly synthesized
Fe-based superconductor Pr[O$_{1-x}$F$_x$]FeAs (F-PrOFeAs) with $T_c=42K$ (the measurement and data
analysis is still in process). All these considerations suggest that the ZBCP observed here is most
possibly related to some type of nodal gap function. In the scanning electron microscopic picture
obtained on the sample surface, it was noted that there are many tiny crystalline stacks distribute
disorderly with finite space between them. Consequently, the tip is easy to penetrate through the
sample surface and rests in a pit, leading to the contacts between the tip and the sides of
surrounded tiny crystals and hence increases the opportunities to detect ZBCP if a nodal pairing
symmetry exists.

\begin{figure}
\onefigure[scale=1.15]{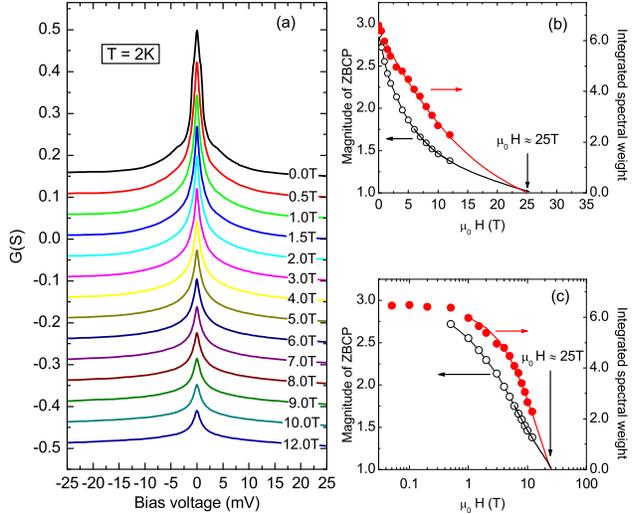} \caption{(a) $G-V$ curves measured in various magnetic fields for
$T=2K$. All the curves except the top one are offset downwards for clarity. (b) Field dependence of
the magnitude of ZBCP and the integrated spectral weight. Both parameters are obtained from the
spectra shown in (a) after normalization by the data of 35K. (c) Same data as that in (b) while the
horizontal axis is in logarithmic scale.} \label{fig:fig3}
\end{figure}

Figure~\ref{fig:fig3}(a) shows the evolution of the ZBCP with increasing field from 0 to 12T at a
fixed temperature of 2K. ZBCP is suppressed continuously whereas survives up to the highest field
in our measurements. We present in Fig.~\ref{fig:fig3}(b) both the height and integrated spectral
weight of the ZBCP. The same data are re-plotted in Fig.~\ref{fig:fig3}(c) while the horizontal
axis is in logarithmic scale. By extrapolating these two parameters to zero independently, we can
obtain consistently the local upper critical field ($H_{c2}$) of 25T from both
Figs.~\ref{fig:fig3}(b) and (c). This value is much smaller than $H_{c2}>50T$ estimated from
resistivity measurements \cite{Chen2008cond,Zhu2008cond}. As mentioned in Ref.\cite{Zhu2008cond},
the upper critical field determined from resistivity reflects mainly the situation of $H\parallel
a-b$ plane since the Cooper pairs within the grains with this configuration will last to the
highest field compared to other grains. Since our point-contact configuration is a local
measurement, it probes the grains with particular orientations, so the determined $H_{c2}$ should
locate between $H_{c2}^{ab}\approx56T$ and $H_{c2}^{c}\approx6T$ by assuming an anisotropy
$\Gamma=H_{c2}^{ab}/H_{c2}^{c}\approx10$ \cite{Zhu2008cond}, in agreement with the result in this
work.

\begin{figure}
\onefigure[scale=1.15]{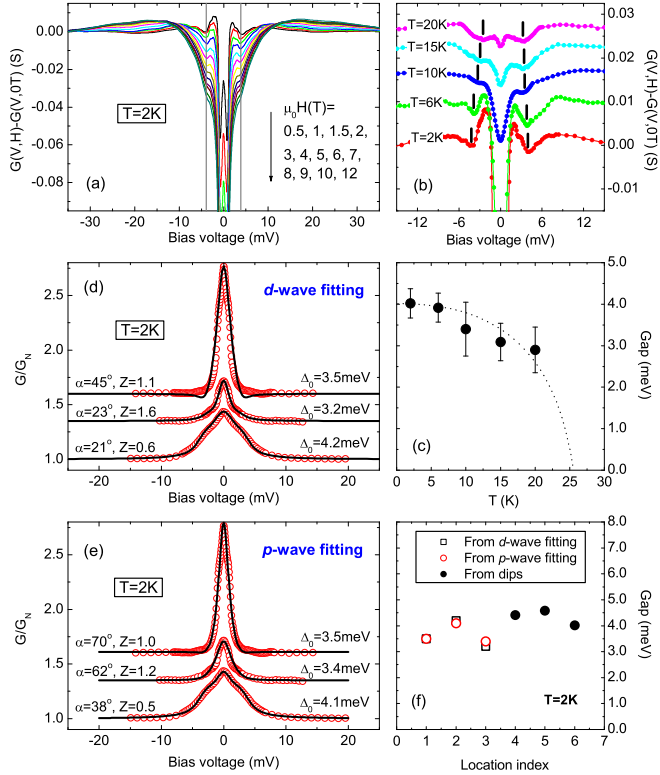} \caption{(a) The conductance of various fields from
Fig.~\ref{fig:fig3}(a) after subtraction of the zero-field conductance. (b) The conductance at low
field of 0.5T or 1.0T after subtraction of the zero-field conductance for various temperatures. The
positions of the dips in each curve are indicated by short bars. (c) Maximum superconducting gap
determined from the data presented in (b). (d) and (e) Fitting the normalized spectra (open
circles) obtained at various locations on the sample surface to the extended BTK model with
$d$-wave and $p$-wave pairing symmetries, respectively. (f) Summing up the maximum gap values of
various locations on the sample surface determined both by dips and fits.} \label{fig:fig4}
\end{figure}

Dagan {\it et al.} \cite{Dagan2000PRB} proposed a convenient method to determine the gap value of a
nodal superconductor by considering that the peak in the density of states at the gap value should
be sensitive to an applied field, while other structures should not be. The method of data analysis
consists simply in subtracting conductances measured in an applied field, from the zero field
conductance. Structures that are not directly related to superconductivity are, in this way,
eliminated \cite{Dagan2000PRB}. When the subtraction procedure is applied to the data of
Fig.~\ref{fig:fig3}(a), as shown in Fig.~\ref{fig:fig4}(a), very clear dips appear symmetrically
around $\pm4$mV and locate at the same bias for fields up to 5T (indicated by two vertical lines).
This is very similar to the case discussed in Ref.\cite{Dagan2000PRB} (except that there is a field
dependent background here), allowing a precise determination of the gap values.
Fig.~\ref{fig:fig4}(b) shows such dip structure for various temperatures and the gap values
determined from these dips are presented in Fig.~\ref{fig:fig4}(c). Since the thermal smearing
effect plays more important role at higher temperatures, the results of higher temperatures shown
in Fig.~\ref{fig:fig4}(c) may be overestimated to some extent. Even though, the decay of the gap
value with increasing temperature is obvious.

Figure~\ref{fig:fig4}(d) shows the estimation of the gap values by fitting the normalized spectra
(measured at different locations) to the extended BTK model with a $d_{x^2-y^2}$ gap function. The
obtained $\Delta_0$ varies in a range of 1mV, indicating a certain degree of inhomogeneity on the
sample surface. Two symmetric slight shoulders have also been observed in some spectra (refer to
Figs.~\ref{fig:fig1}(a) and (c)), which can not be fitted with the single gap model used here. This
may be related to another gap opening on other Fermi surfaces though it contributes little to the
measured spectra. It should be mentioned that our data support a nodal gap function of F-LaOFeAs
while can not distinguish between $d$-wave and $p$-wave. Since the superconducting phase of
F-LaOFeAs is located close to the ferromagnetic phase \cite{Kamihara2008JACS}, it is possible that
spin triplet pairing, such as $p$-wave pairing symmetry, is favored in this material. In fact, very
similar ZBCP was observed in Sr$_2$RuO$_4$ single crystal with spin triplet superconductivity
\cite{Mao2001PRL}. Fig.~\ref{fig:fig4}(e) shows the calculations with a $p$-wave gap function of
$\Delta=\Delta_0sin\theta$, which is consistent with the experimental data as well as the $d$-wave
fits. In both approaches, the nodal characteristic and the maximum value of the gap function of
F-LaOFeAs can be confirmed consistently. Most recently, NMR data present strong evidence of singlet
pairing in PrFeAsO$_{0.89}$F$_{0.11}$ \cite{NMR_ZhengGQ2008cond}, thus the singlet pairing symmetry
such as $d$-wave seems to be more possible than the $p$-wave pairing for the sample studied here.
In Fig.~\ref{fig:fig4}(f), we can see that the gap values determined both by fitting procedure and
from the dip structure are well consistent with each other, which can be summed up as
$\Delta_0=3.9\pm0.7$meV. This result is in good agreement with $\Delta_0=3.4\pm0.5$meV from our
specific heat measurement \cite{Mu2008cond} and $\Delta_0=4.0\pm0.6$meV from our $H_{c1}$
measurement \cite{RenC2008cond} on the same material. Thus, we can calculate the
Bardeen-Cooper-Schrieffer coherence length $\xi_{BCS}=\hbar\nu_F/\pi\Delta_0$ where the in-plane
Fermi velocity $\nu_F^{ab}=1.30\times10^7$cm/s \cite{Singh2008cond}. Using $H_{c2}^{c}\approx6T$
mentioned above \cite{Zhu2008cond}, we can calculate the Ginzburg-Landau coherence length
$\xi_{GL}^{ab}=\sqrt{\phi_0/2\pi H_{c2}^{c}}$ with $\phi_0$ the flux quantum. The obtained
$\xi_{BCS}=70{\AA}$ is in good agreement with the calculated $\xi_{GL}^{ab}=74{\AA}$ within the
experimental errors. We also estimated the mean free path $l\approx100{\AA}$ from resistivity by
using $k_Fl=hc/e^2\rho$ (where $k_F\simeq m\nu_F/\hbar$, $c=8.7{\AA}$ is the lattice constant along
the normal to the FeAs plane, and $h/e^2=26$K$\Omega$ is the quantum resistance). These estimations
self-consistently suggest that F-LaOFeAs is in the moderate clean limit, which is different from
the dirty-limit case of the electron-doped cuprates with similar $T_c$ values \cite{Dagan2007PRL}.
It is also the reason why a clear ZBCP can be observed in this work and the $\Delta_0$ can be
derived properly from the specific heat data\cite{Mu2008cond} for F-LaOFeAs.

\section{Conclusion}
In summary, we have studied point-contact spectroscopy of the junctions built up between a
normal-metal tip and the newly discovered Fe-based layered superconductor
LaO$_{0.9}$F$_{0.1-\delta}$FeAs. A distinct zero-bias conductance peak was observed and can be
related to the surface Andreev bound states. Our data present a possible evidence that
LaO$_{0.9}$F$_{0.1-\delta}$FeAs is a nodal superconductor with a maximum gap of $3.9\pm0.7$meV,
which is in good agreement with our recent specific heat and $H_{c1}$ data.


\acknowledgments This work is supported by the Natural Science Foundation of China, the Ministry of
Science and Technology of China ( 973 project No: 2006CB601000, 2006CB921802, 2006CB921300 ), and
Chinese Academy of Sciences (Project ITSNEM).

\end{document}